\documentclass[11pt, a4paper]{article}

\usepackage{latexsym}
\usepackage{graphicx}
\usepackage{cite}
\usepackage{amsmath}
\usepackage{amssymb}
\begin{document}
\title{Ergodicity and Measurements In Static and Dynamic Light Scattering}
\author{Yong Sun{\footnote{Email: ysun200611@yahoo.ca}}\\
\emph{ysun200611@yahoo.ca}
\\\emph{Burnaby, BC, Canada}}
\maketitle

\begin{abstract}

Static Light Scattering (SLS) and Dynamic Light Scattering (DLS)
are very important techniques to study the characteristics of 
nano-particles in dispersion. The data of SLS is determined by the optical characteristic
and the measured values of DLS are determined by optical 
and hydrodynamic characteristics of different size nano-particles.
In general, the nano-particles investigated are considered to be cross-linked soft particles or  
hyper-branched chains with 
a three dimensional network structure. Therefore the density of these nano-particles is not homogeneous and
the different parts have different optical characteristics. However 
our experiments reveal that the long time average data of scattered intensity can be perfectly described by homogeneous spherical model. 
Based on the size distribution obtained using the SLS or TEM technique and the 
relation between the static and hydrodynamic radii, all the calculated and measured values of 
$g^{\left( 2\right) }\left( \tau \right) $ investigated are also consistent very well. Since the
long time average scattered intensity of non-homogeneous spherical nano-particles can be perfectly described by homogeneous spherical model,
the phenomenons of ergodicity happen in our experiments. The results also reveal that the root mean-square radius of gyration $\left\langle R_{g}^{2}\right\rangle ^{1/2}$ obtained using the Zimm plot, Berry plot or Guinier
plot is an optical weight size. Due to the different optical average methods of the root mean-square radius of gyration $\left\langle R_{g}^{2}\right\rangle ^{1/2}$ and apparent hydrodynamic radius  $R_{app,h}$ and the complex hydrodynamic structures, the dimensionless shape parameter $\rho =\left\langle R_{g}^{2}\right\rangle ^{1/2}/R_{app,h}$ has lost the physical significance to judge the shapes of nano-particles.

\end{abstract}

\section{INTRODUCTION}

Light Scattering has been considered to be a well established
technique and applied in physics, chemistry, biology, etc. as an
essential tool to investigate the characteristics of nano-particles
in dispersion. It consists of two parts: Static Light Scattering (SLS)
and Dynamic Light Scattering (DLS) techniques. The SLS technique measures the relation between
the scattering angles and scattered light intensity and then
simplifies the relation to the Zimm plot, Berry plot or Guinier
plot to get the root mean-square radius of gyration
$\left\langle R_{g}^{2}\right\rangle ^{1/2}$ provided that the particle sizes are small\cite{re1,re2}. 
$\left\langle R_{g}^{2}\right\rangle ^{1/2}$
has been considered to be determined by the shapes of nano-particles.
For example, under a mono-disperse model, the dimensionless shape parameter 
$\rho =\left\langle R_{g}^{2}\right\rangle ^{1/2}/R_h$  is considered to be about 0.775
for homogeneous spherical particles and 1.505 for linear chains. However a lot of nano-particles are
made using chemical reactions. They consist of different atoms such as hydrogen, carbon, oxygen and nitrogen etc.
When these atoms form cross-linked soft particles or  
hyper-branched chains with a three dimensional network structure dispersed in solvent, these nano-particles can not 
be considered as homogeneous spherical nano-particles whenever the density or optical characteristic are considered.

\noindent For many dilute  spherical nano-particles, one method\cite{re3} is proposed to measure
the size distribution of nano-particles accurately. The size 
distribution of nano-particles in dispersion is chosen to be a Gaussian distribution and the electro-magnetic characteristics has been considered to be homogeneous.
Using a non-linear least square fitting program, 
the mean static radius $\left\langle R_{s}\right\rangle $ and standard deviation $\sigma$
are measured accurately. For PNIPAM samples, the long time average scattered light intensity investigated
are consistent with the expected values of the homogeneous spherical particle model and for the polystyrene spherical particles investigated, the measured values of $\left\langle R_{g}^{2}\right\rangle ^{1/2}$ are consistent with these values 
calculated based on the particle size information measured using the Transmission Electron Microscopy (TEM) technique provided by the supplier very well, respectively. 
Using the size distribution obtained using the SLS or TEM technique, with the relation between the static and dynamic radii, the calculated results and measured data of $g^{\left( 2\right)}\left( \tau \right)$ at all the scattering angles investigated are consistently perfect. For the PNIPAM samples, the fitting results obtained using the non-linear least square fitting program  not only are
consistent with the data of scattered light intensity very well but also the residuals show random.

\section{THEORY}

The nano-particles in dispersion are considered to be homogeneous spherical particles with polarizability $\alpha$.
When homogeneous spherical particles are considered and Rayleigh-Gans-Debye(RGD) approximation is valid, the following equation 
of the average scattered light intensity of a dilute non-interacting particle system in
unit volume can be obtained for vertically incident polarized light. 

\begin{equation}
\frac{I_{s}}{I_{inc}}=\frac{4\pi ^{2}\sin ^{2}\theta
_{1}n_{s}^{2}\left( \frac{dn}{dc}\right) _{c=0}^{2}c}{\lambda
^{4}r^{2}}\frac{4\pi \rho }{3} \frac{\int_{0}^{\infty
}R_{s}^{6}P\left( q,R_{s}\right) G\left( R_{s}\right)
dR_{s}}{\int_{0}^{\infty }R_{s}^{3}G\left( R_{s}\right) dR_{s}},
\label{mainfit}
\end{equation}
where $I_{inc}$ is the incident light intensity, $ I_{s}$ is the intensity of the scattered light that reaches the detector, $R_{s}$ is the static radius of a particle, $\ q=\frac{4\pi
}{\lambda }n_{s}\sin \frac{\theta }{2}$ is the scattering vector,
$\lambda $ is the wavelength of incident light in vacuo, $n_{s}$\
is the solvent refractive index, $ \theta $ is the scattering
angle, $\rho $ is the density of particles, $r$
is the distance between the scattering particle and the point of
intensity measurement, $c$ is the mass concentration of particles, 
$\theta _{1}$ is the angle between the polarization of the
incident electric field and the propagation direction of the
scattered field, 
$P\left( q,R_{s}\right) $ is the form factor of homogeneous
spherical particles

\begin{equation}
P\left( q,R_{s}\right) =\frac{9}{q^{6}R_{s}^{6}}\left( \sin \left(
qR_{s}\right) -qR_{s}\cos \left( qR_{s}\right) \right) ^{2}
\label{P(qr)}
\end{equation}
and $G\left( R_{s}\right) $ is the number distribution of
particles in dispersion which is chosen as a
Gaussian distribution

\begin{equation}
G\left( R_{s};\left\langle R_{s}\right\rangle ,\sigma \right)
=\frac{1}{ \sigma \sqrt{2\pi }}\exp \left( -\frac{1}{2}\left(
\frac{R_{s}-\left\langle R_{s}\right\rangle }{\sigma }\right)
^{2}\right) ,
\end{equation}
where $\left\langle R_{s}\right\rangle $ is the mean static radius
and $\sigma $ is the standard deviation.

\noindent Based on the particle size information obtained using the SLS technique, 
for dilute homogeneous spherical particles, $g^{\left(1\right)}(\tau )$ is

\begin{equation}
g^{\left( 1\right) }\left( \tau \right) =\frac{\int R_{s}^{6}
P\left( q,R_{s}\right)G\left( R_{s}\right) \exp \left( -q^{2}D\tau
\right) dR_{s}}{\int R_{s}^{6}P\left( q,R_{s}\right) G\left(
R_{s}\right) dR_{s}}, \label{Grhrs}
\end{equation}

\noindent Here the Stokes-Einstein relation is still considered to be 
true for nano-particles,

\begin{equation}
D=\frac{k_{B}T}{6\pi \eta _{0}R_{h}},
\end{equation}
where $\eta _{0}$, $k_{B}$ and $T$ are the viscosity of solvent,
Boltzmann's constant and absolute temperature respectively, then the
hydrodynamic radius $R_{h}$ can be obtained.

\noindent The relation between the static and
hydrodynamic radii in this work is assumed to be
\begin{equation}
R_{h}=kR_{s},  \label{RsRh}
\end{equation}
where $k$ is a constant. Based on the Siegert relation between
$g^{\left( 2\right) }\left( \tau \right) $ and $g^{\left( 1\right)
}\left( \tau \right)$ \cite{re4}

\begin{equation}
g^{\left( 2\right) }\left( \tau \right) =1+\beta \left( g^{\left(
1\right) }\right) ^{2},  \label{G1G2}
\end{equation}
the function between the SLS and DLS techniques is built and the values of $
g^{\left( 2\right) }\left( \tau \right) $ can be expected based on
the particle size information measured using the SLS technique.

\section{RESULTS AND DISCUSSION}

\subsection{PNIPAM samples}

All the PINPAM microgel samples are made using the PNIPAM monomers and cross-linker $N,N'$-methylenebiscylamide.
The potassium was introduced to initiate polymerization. When the PNIPAM microgel samples are made, they are 
considered to have three dimensional structures with
cross-link networks or hyper-branched chains\cite{re5}. Therefore if the density or optical properties are considered,
these PNIPAM microgels can not have homogeneous characteristics. However, when the long time average intensity of
scattered light was recorded and investigated, they reveal that these nano-particles investigated occupy strong characteristic of homogeneous spherical microgels.

Table \ref{table3} shows the results of the data of the PNIPAM-1 measured using the SLS technique
at a temperature of 302.33 K was fitted using Eq. \ref{mainfit}. The results reveal that the fitting results are 
determined by the fitting ranges. If the fitting range is large enough, uncertainties in the parameters 
$\left\langle R_{s}\right\rangle $ and $\sigma $ are small and the fitting values of $\left\langle R_{s}\right\rangle $ and $\sigma $ stabilize. If the fitting range gets larger, the values of $\left\langle
R_{s}\right\rangle $ and $\sigma $ begin to lose stability and $\chi ^{2}$
grows. Therefore the stable
fitting results $\left\langle R_s\right\rangle = 254.3 \pm 0.1$ nm and
$\sigma = 21.5 \pm 0.3$ nm obtained in the scattering vector range
between 0.00345 and 0.01517 nm$^{-1}$ are chosen as the particle size
information measured using the SLS technique. Fig. \ref{figPNI-302fit} shows all the experimental data,
stable fitting results and residuals in the scattering vector range between 0.00345 and
0.01517 nm$^{-1}$ together. The figure reveals that the fitting results and measured data are consistent very well and the residuals are random. That means 
the homogeneous spherical model can describe the long time average scattered intensity perfectly.

\begin{center}
\begin{tabular}{|c|c|c|c|}
\hline $q$ ($10^{-3}$ nm$^{-1}$) & $\left\langle
R_{s}\right\rangle$ (nm) & $ \sigma$ (nm) & $\chi ^{2}$ \\
\hline 3.45 to 9.05 & 260.09$\pm $9.81 & 12.66$\pm $19.81 & 1.64 \\
\hline 3.45 to 11.18 & 260.30$\pm $1.49 & 12.30$\pm $3.37 & 1.65 \\
\hline 3.45 to 13.23 & 253.45$\pm $0.69 & 22.80$\pm $0.94 & 2.26 \\
\hline 3.45 to 14.21 & 254.10$\pm $0.15 & 21.94$\pm $0.36 & 2.03 \\
\hline 3.45 to 15.17 & 254.34$\pm $0.12 & 21.47$\pm $0.33 & 2.15 \\
\hline 3.45 to 17.00 & 255.40$\pm $0.10 & 17.32$\pm $0.22 & 11.02 \\
\hline
\end{tabular}
 \makeatletter\def\@captype{table}\makeatother
\caption{The results obtained using Eq. \ref{mainfit} for
PNIPAM-1 at different scattering vector ranges and a temperature
of 302.33 K.}\label{table3}
\end{center}

\begin{center}
   \includegraphics[width=0.35\textwidth,angle=-90]{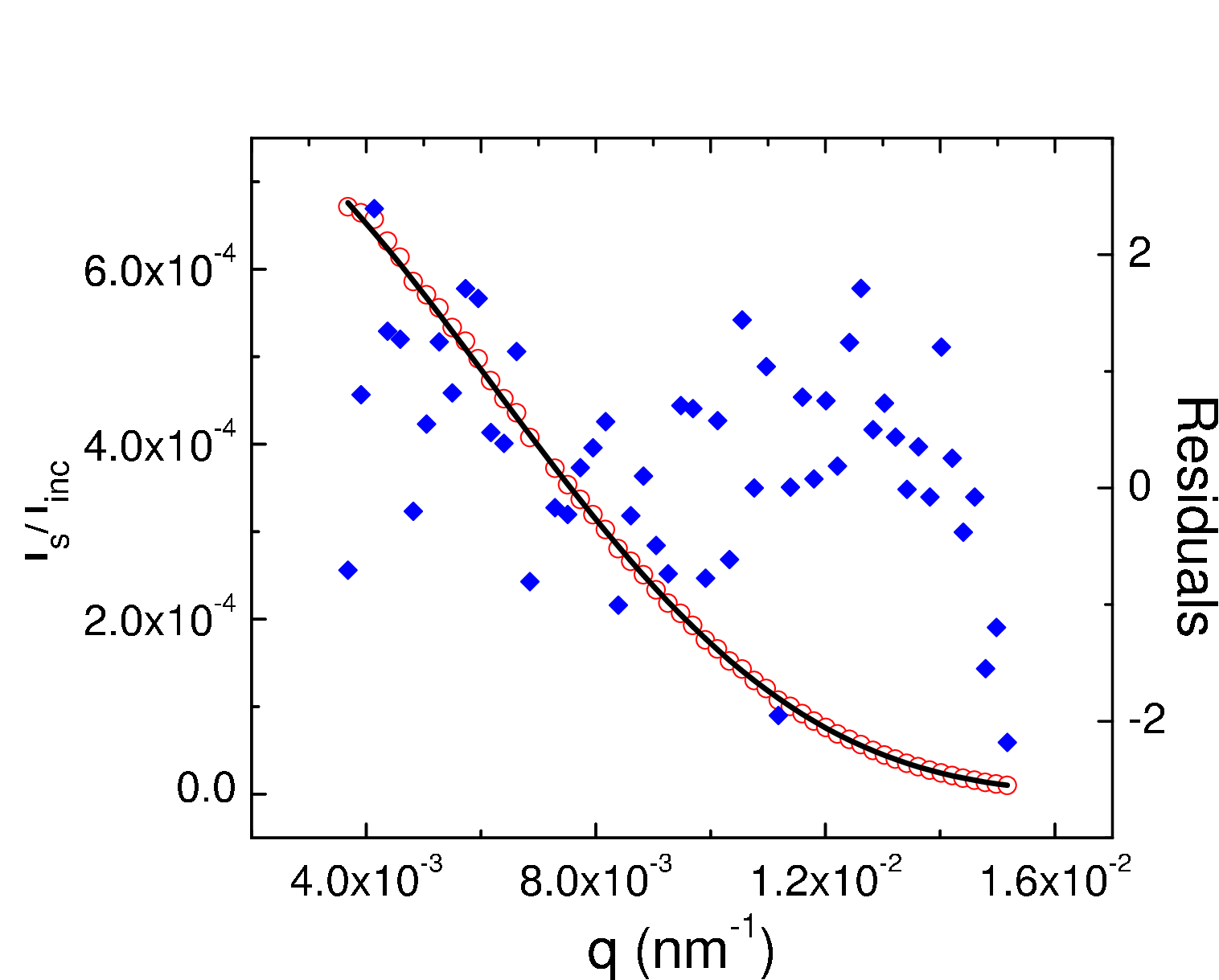}
      \makeatletter\def\@captype{figure}\makeatother
   \caption{The experimental data and stable fitting results for PNIPAM-1 at a temperature of 302.33 K.
   The circles show the experimental data, the line shows the fitting results and the diamonds show the
residuals: $\left( y_{i}-y_{fit}\right) /\sigma _{i}$.}
 \label{figPNI-302fit}
\end{center}

\noindent  Since the PNIPAM microgel samples have complex structures, the hydrodynamic characteristic 
will be affected significantly. The relation between the static and hydrodynamic radii will also change. If the constant $k$ in Eq. \ref{RsRh} is equal to 1.21, all the values of $g^{\left( 2\right) }\left( \tau \right) $ at the scattering
angles 30$^\mathrm o$, 50$^\mathrm o$ and 70$^\mathrm o$ can be calculated based on the particle size information measured using the SLS technique. Fig. \ref{figPNI-302cal} shows all the measured data and calculated values.
Fig. \ref{figPNI-302cal} reveals that
the calculated values and experimental data are  consistent. This means that the long time average data of dynamic light also satisfies the homogeneous spherical condition very well.

\begin{center}
  \includegraphics[width=0.35\textwidth,angle=-90]{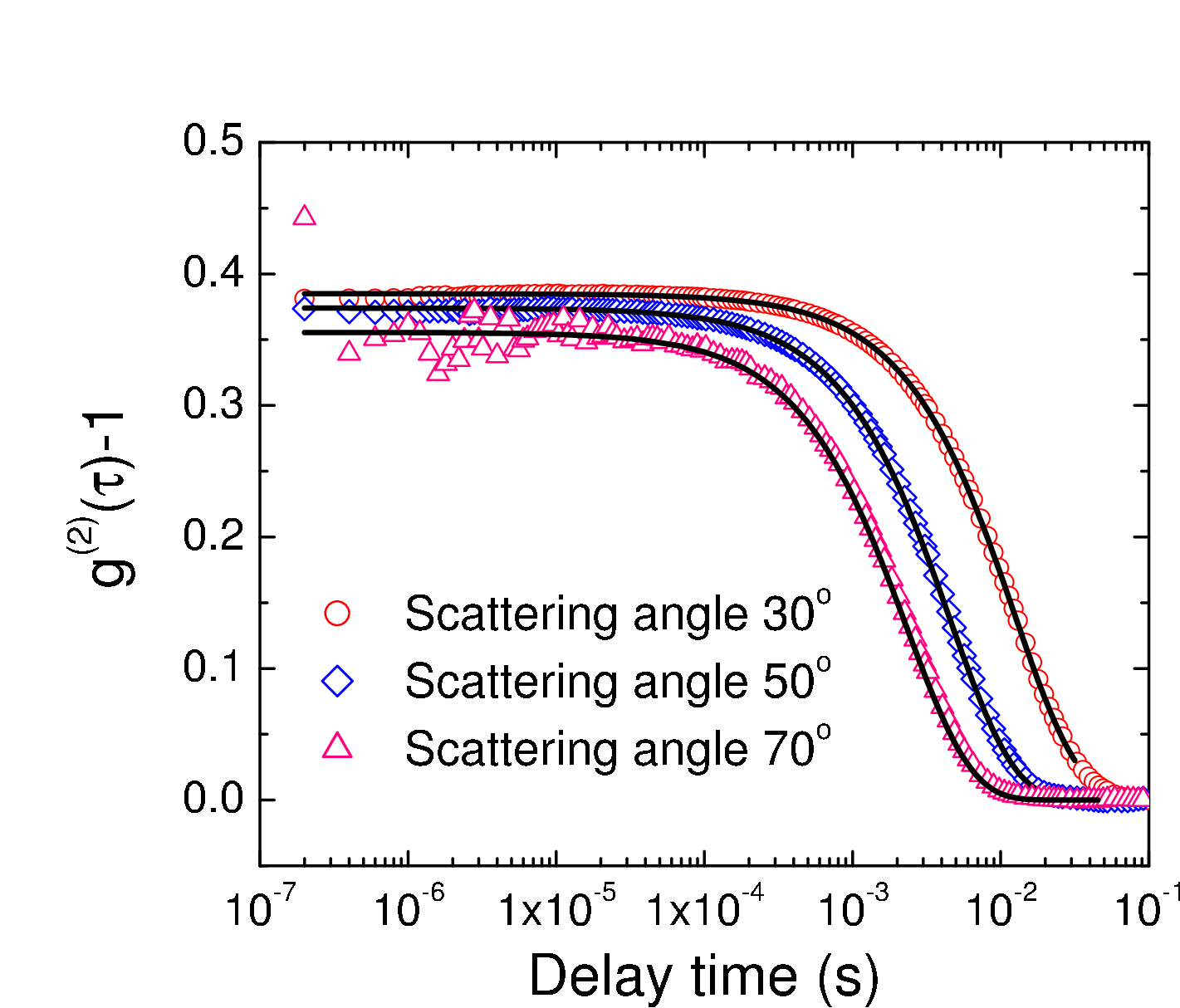}
   \makeatletter\def\@captype{figure}\makeatother
  \caption{The measured data and expected values of $g^{\left(
2\right) }\left( \tau \right)$ of PNIPAM-1 at a temperature of
302.33 K. The symbols represent the experimental data and the lines
show the expected values calculated under $R_{h}=1.21R_{s}$.}
\label{figPNI-302cal}
\end{center}

\noindent  When the PNIPAM samples are measured at high temperatures, all the fitting situations using
Eq. \ref{mainfit} are the same as that of PNIPAM-1 at a temperature
of 302.33 K. The results are 
determined by the fitting ranges. If the fitting range is large enough, uncertainties in the parameters 
$\left\langle R_{s}\right\rangle $ and $\sigma $ are small and the fitting values of $\left\langle R_{s}\right\rangle $ and $\sigma $ stabilize.  For the PNIPAM-5 sample at a temperature of
312.66 K,  the stable fitting results $\left\langle R_s\right\rangle = 139.3 \pm 0.3$ nm and
$\sigma = 12.4 \pm 0.6$ nm are chosen as the particle size information measured using the SLS
technique obtained in the scattering vector range
between 0.00345 and 0.02555 nm$^{-1}$. The stable fitting results and
residuals are shown in Figure \ref{figPNIhT}a. The figure reveals that the fitting results and measured data are consistent and the residuals are random. That means 
the homogeneous spherical model can describe the long time average scattered intensity perfectly. 
If the constant $k$ in Eq. \ref{RsRh} is equal to 1.1, all the values of $g^{\left( 2\right) }\left( \tau \right) $ at the scattering
angles 30$^\mathrm o$, 50$^\mathrm o$, 70$^\mathrm o$ and 100$^\mathrm o$ are calculated based on the particle size information measured using the SLS technique. Fig. \ref{figPNIhT}b shows all the measured data and calculated values
of $g^{\left( 2\right) }\left( \tau \right) $.
Fig. \ref{figPNIhT}b also reveals that
the calculated values and experimental data are  consistent very well. This also means that the long time average data of dynamic light scattering also satisfies the homogeneous spherical condition very well.

\begin{center}
  $\begin{array}{c@{\hspace{0in}}c}
     \includegraphics[width=0.35\textwidth,angle=-90]{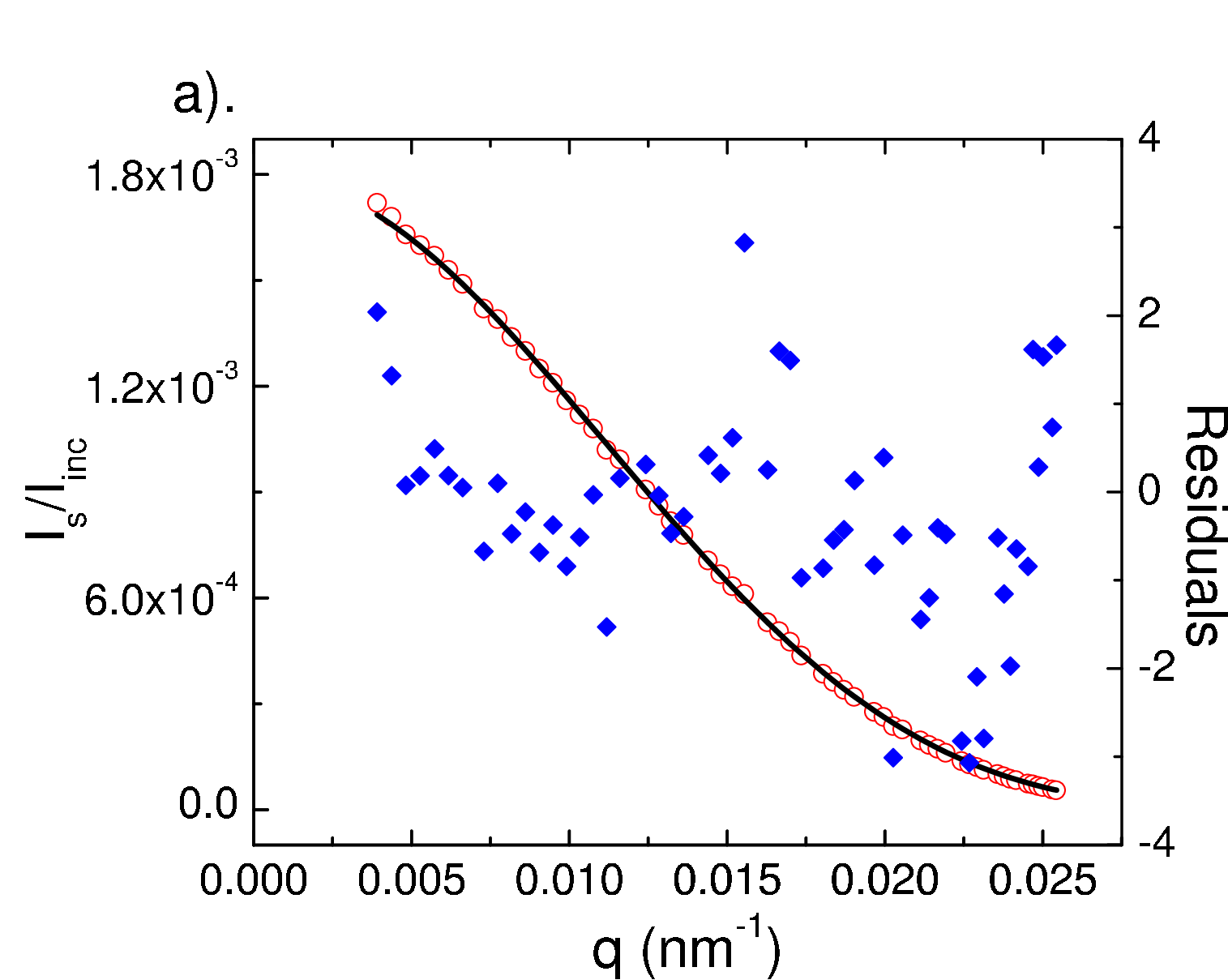} &
     \includegraphics[width=0.35\textwidth,angle=-90]{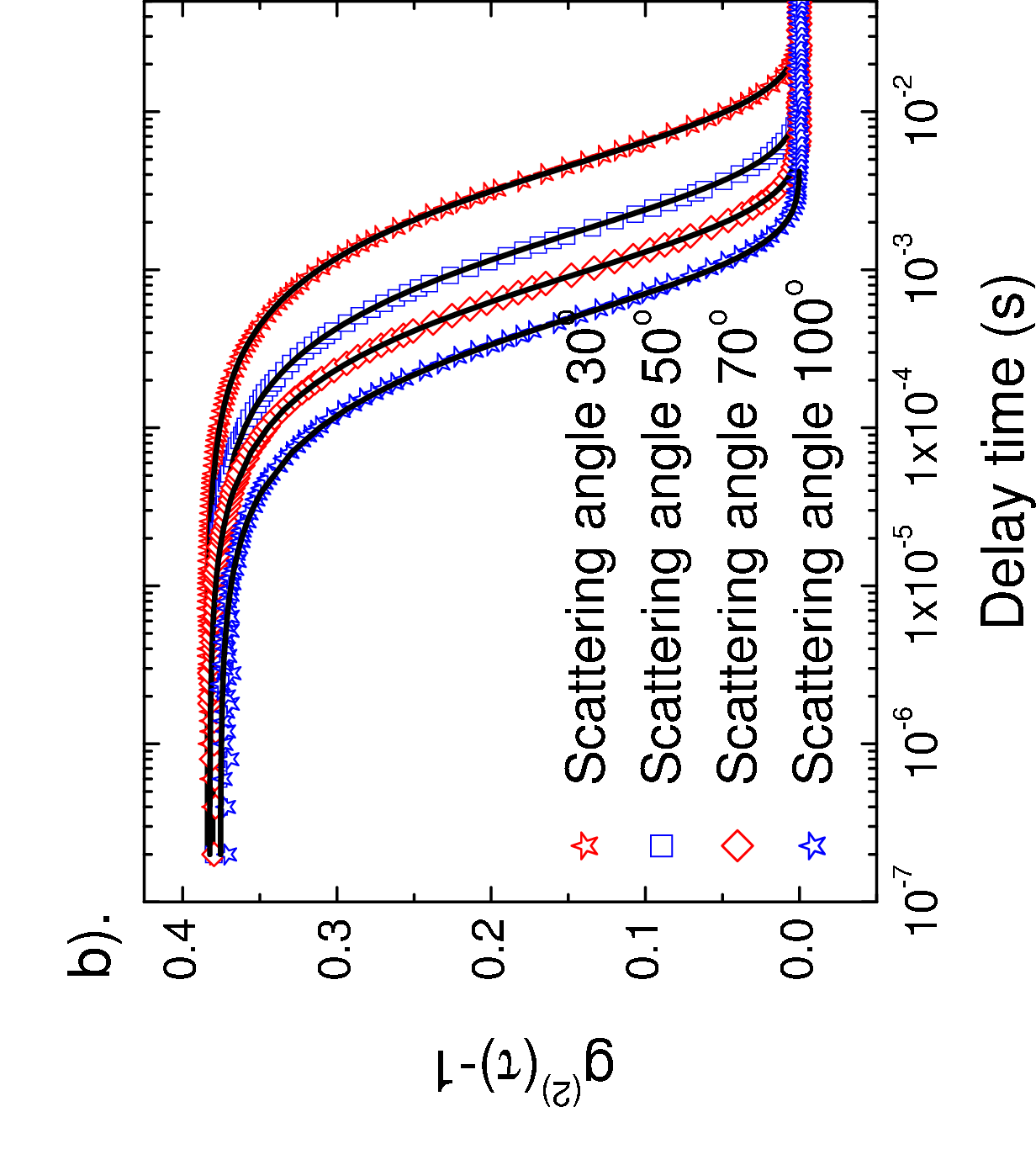} \\ [0.0cm]
    \end{array}$
   \end{center}\vspace{-0.5cm}
 \makeatletter\def\@captype{figure}\makeatother
  \caption{The static and dynamic results of PNIPAM-5 at a temperature 312.66 K. 
  a). The measured data and stable fitting results. The circles represent the experimental data, the line shows the fitting results and the diamonds represent the
residuals: $\left( y_{i}-y_{fit}\right) /\sigma _{i}$. b). The
measured data and expected values of $g^{\left( 2\right)
}\left( \tau \right)$. The symbols represent the measured data and
the lines show the expected values calculated under
$R_{h}=1.1R_{s}$.} \label{figPNIhT}

\noindent  For all other PNIPAM samples investigated, the situation are the same. The fitting results are 
determined by the fitting ranges. If the fitting range is large enough, uncertainties in the parameters 
$\left\langle R_{s}\right\rangle $ and $\sigma $ are small and the fitted values of $\left\langle R_{s}\right\rangle $ and $\sigma $ stabilize. The calculated values of $g^{\left( 2\right) }\left( \tau \right) $
obtained using the particle size distribution
measured by the SLS technique and measured data are consistent very well.
All the results reveal that the long time average data of static and dynamic light scattering can be described by the homogeneous spherical model very well.

\subsection{Standard polystyrene latex samples}

In this section, the commercial nano-particles are used to test the homogeneous spherical model. The commercial
particle size information was measured using the TEM technique provided by the supplier. Based on the calculation
using the homogeneous model and commercial data, the ``phase shift'' $\frac{4\pi
}{\lambda }R|m-1|$ \cite{re6, re7} of  Latex-1
and Latex-2 are 0.13 and 0.21 respectively. Since they do not exactly satisfy the rough
criterion that a RGD approximation\cite{re6} is valid, 
the mono-disperse model $G\left( R_{s}\right) =\delta \left(
R_{s}-\left\langle R_{s}\right\rangle \right) $ was used to
fit the static data to obtain the values of $\left\langle R_{s}\right\rangle $.

Table \ref{table2} listed all the results measured using the
SLS and TEM techniques.  Figs. \ref{figPolyfitcal}a and \ref{figPolyfitcal}b are also show all the measured and fitting
results for two polystyrene latex samples respectively. All the results reveal that the homogeneous spherical model still
can describe the scattered light intensity very well.

\begin{table}[ht]
\begin{center}
\begin{tabular}{|c|c|c|}
\hline $\left\langle R\right\rangle_{comm}$ (nm) & $\sigma_{comm}$
(nm) & $\left\langle R_{s}\right\rangle$ (nm)\\
\hline 33.5(Latex-1) & 2.5 & 33.3$\pm $0.2 \\
\hline 55(Latex-2) & 2.5 & 56.77$\pm $0.04 \\
\hline
\end{tabular}
\caption[]{Size information measured using the TEM and SLS
techniques.}\label{table2}
\end{center}
\end{table}

\begin{center}
  $\begin{array}{c@{\hspace{0in}}c}
     \includegraphics[width=0.35\textwidth,angle=-90]{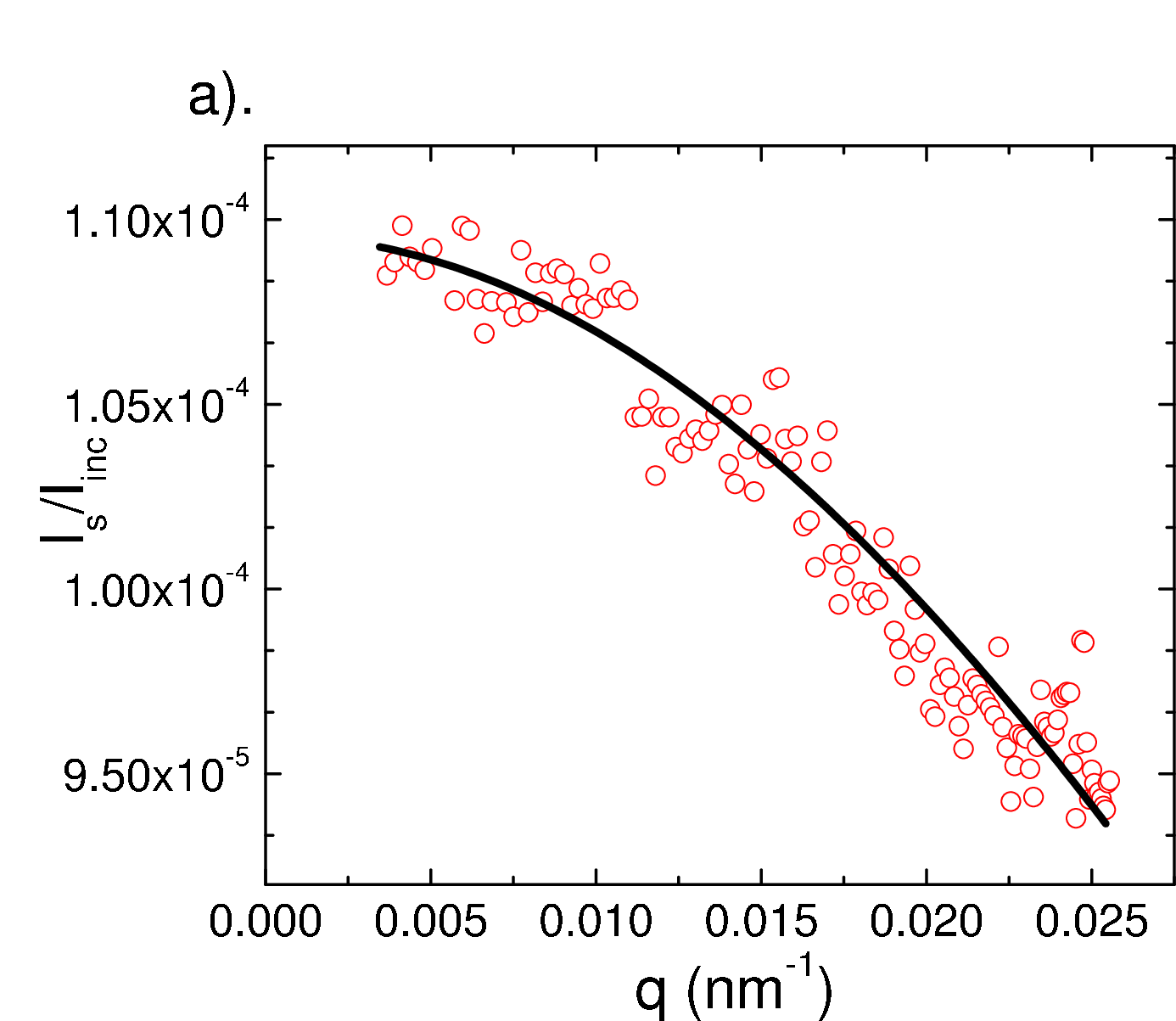} &
     \includegraphics[width=0.35\textwidth,angle=-90]{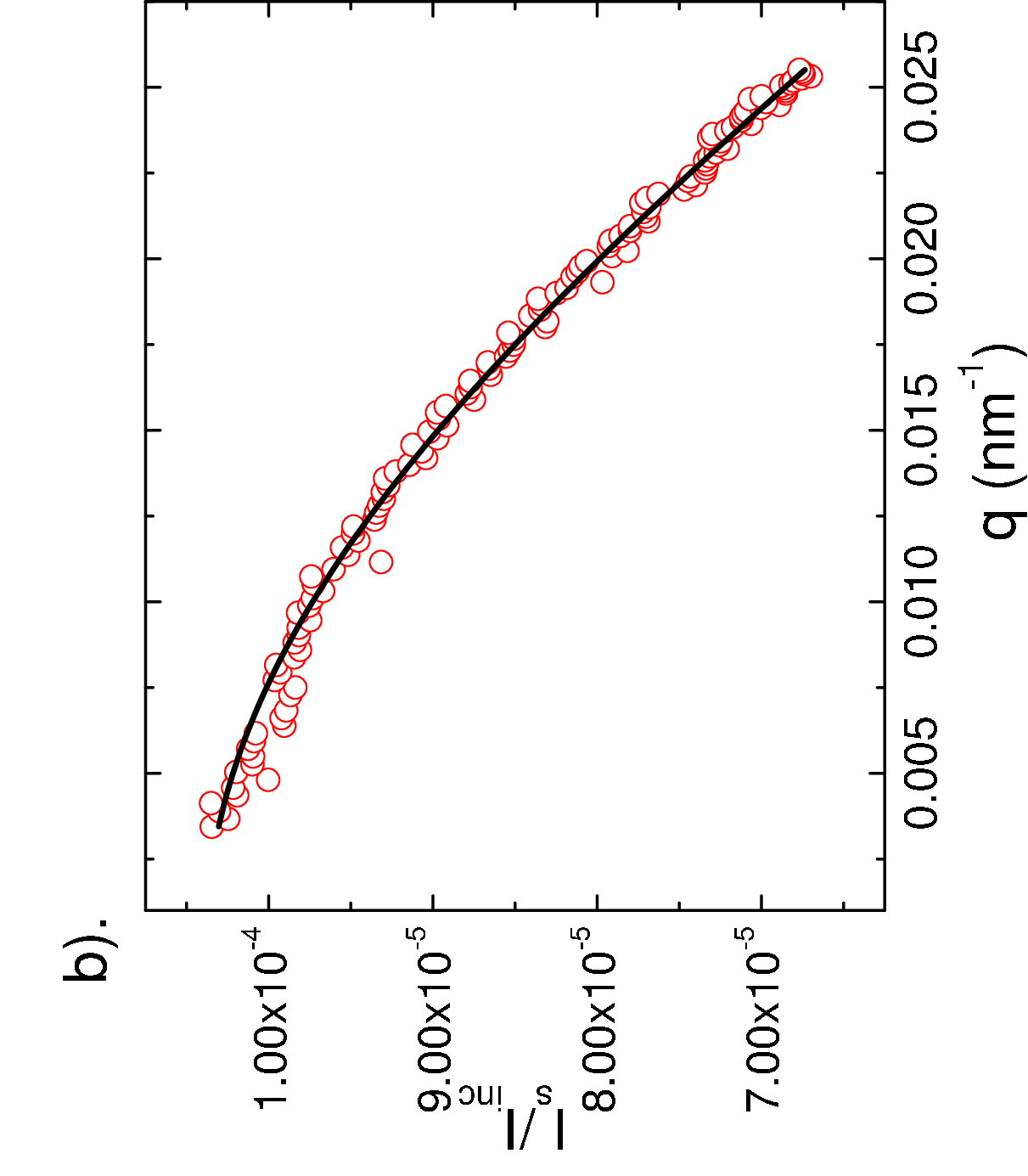} \\ [0.0cm]
    \end{array}$
   \end{center}\vspace{-0.5cm}
 \makeatletter\def\@captype{figure}\makeatother
\caption[] {a). The experimental data and fitting results of
$I_{s}/I_{inc}$ for Latex-1 and b). Latex-2. The circles 
show the experimental data and the line
represents the fitting results of $I_{s}/I_{inc}$.} \label{figPolyfitcal}

\noindent  Next, based on the commercial particle size information, all the values of 
$g^{\left( 2\right) }\left( \tau \right) $ were calculated at the scattering
angles 30$^\mathrm o$, 60$^\mathrm o$, 90$^\mathrm o$,
120$^\mathrm o$ and 150$^\mathrm o$ and a temperature of 298.45 K
for Latex-1, 298.17 K for Latex-2 respectively. During these calculations,
the constant $k$ in Eq. \ref{RsRh} is set to be 1.1 for
Latex-1 and 1.2 for Latex-2, respectively. All the measured and calculated data are shown 
in Figs. \ref{figPolyDLScal}a and \ref{figPolyDLScal}b, respectively.
Figure \ref{figPolyDLScal} reveals that the homogeneous spherical model still
can descript the measured data of $g^{\left( 2\right) }\left( \tau \right) $ very well.

\begin{center}
  $\begin{array}{c@{\hspace{0in}}c}
     \includegraphics[width=0.35\textwidth,angle=-90]{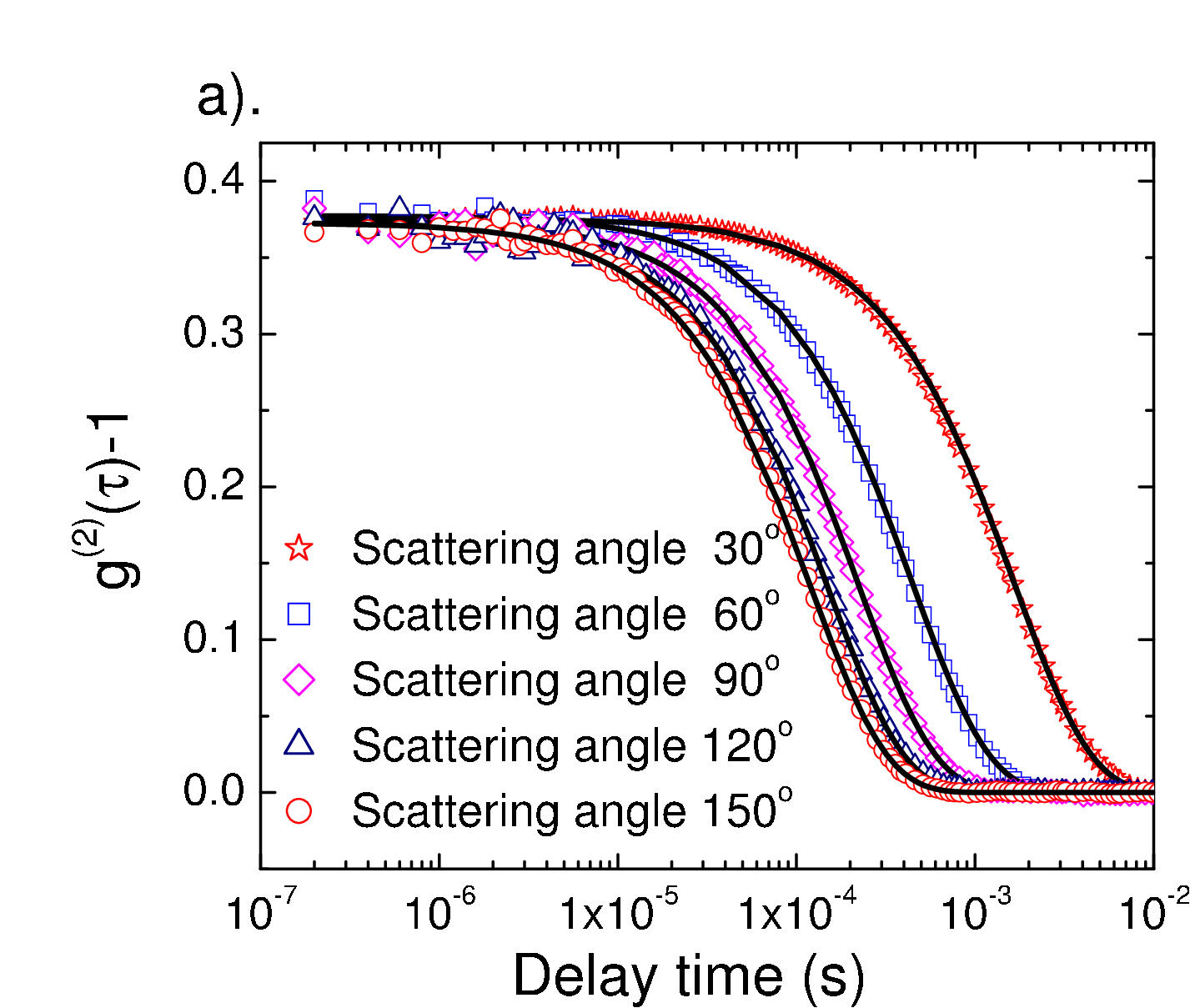} &
     \includegraphics[width=0.35\textwidth,angle=-90]{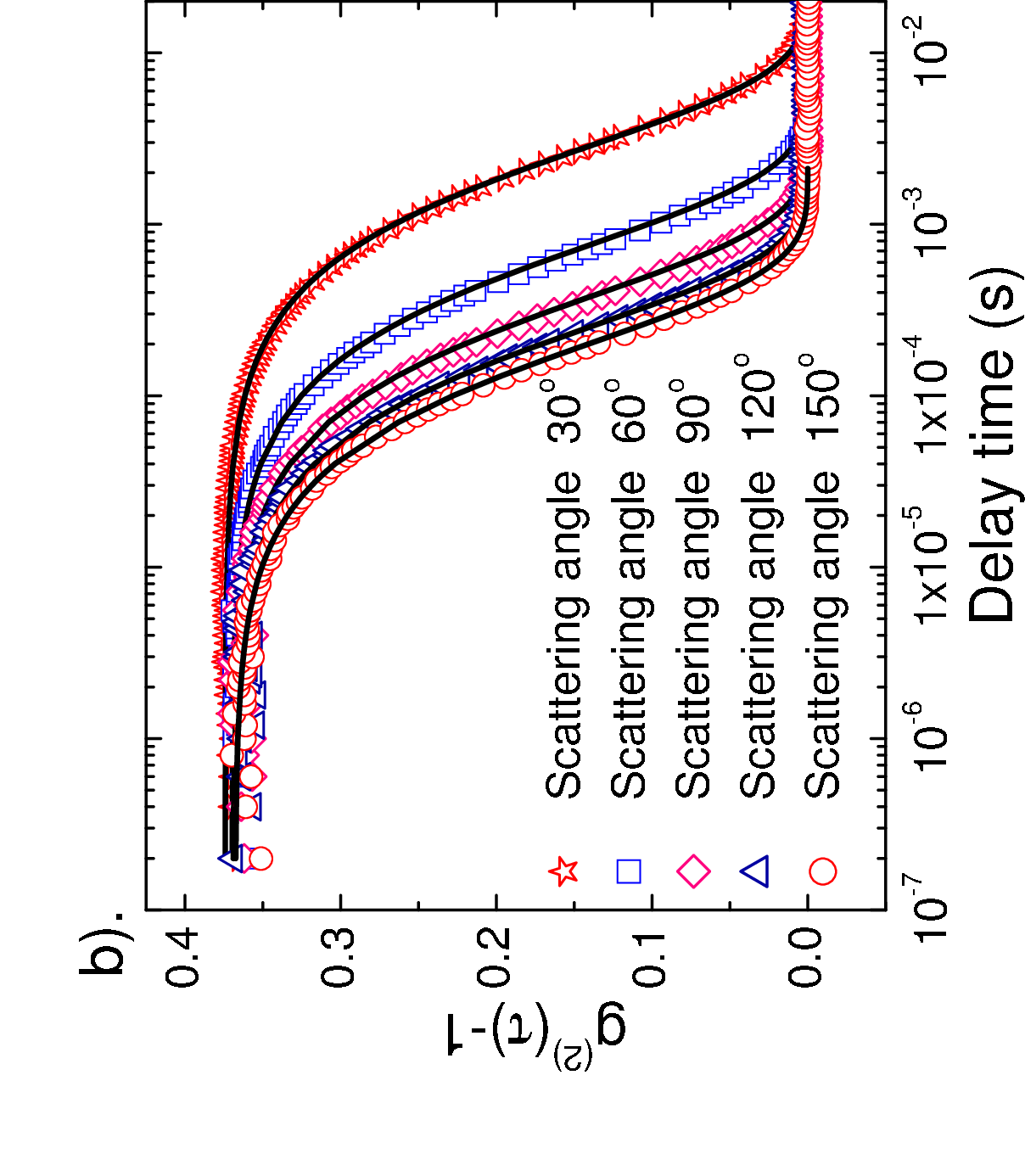} \\ [0.0cm]
    \end{array}$
   \end{center}\vspace{-0.5cm}
 \makeatletter\def\@captype{figure}\makeatother
  \caption{The experimental data and expected values of $g^{\left(
2\right) }\left( \tau \right) $. a). Latex-1 and b).
Latex-2. The symbols represent the experimental data and the lines show
the expected values calculated under $R_{h}=kR_{s}$.}
\label{figPolyDLScal}

\noindent  Furthermore the root mean square radius of gyration 
${\left\langle {R_g}^2\right\rangle}^{1/2} $ also can be used to test the homogeneous spherical model.
The values of ${\left\langle {R_g}^2\right\rangle}^{1/2} $ can be calculated using the commercial
particle size information. Meanwhile they also can be measured using the Zimm plot analysis. 
All these results are listed in Table \ref{table7}. That all the results are consistent respectively reveals that the homogeneous spherical model still can describe the scattered light intensity very well.

\begin{center}
\begin{tabular}{|c|c|c|}
\hline Sample & ${\left\langle {R_g}^2\right\rangle}^{1/2}_{cal} $ & ${\left\langle 
{R_g}^2\right\rangle}^{1/2}_{Zimm} $\\
\hline 33.5(nm) & 26.9  & 26.9$\pm $0.5 \\
\hline 55 (nm)  & 43.24 & 46.8$\pm $0.3  \\
\hline 90 (nm) & 70.1   & 69.0$\pm $2.0 \\
\hline
\end{tabular}
 \makeatletter\def\@captype{table}\makeatother
\caption{Values of ${\left\langle {R_g}^2\right\rangle}^{1/2}_{cal} $ and
 ${\left\langle {R_g}^2\right\rangle}^{1/2}_{Zimm} $.}
\label{table7}
\end{center}

\noindent The results also reveal that the root mean square radius of gyration ${\left\langle {R_g}^2\right\rangle}^{1/2} $ is an optical weight size. The size information is obtained from the long time average scattered light intensity and the scattered light is determined by the optical characteristics of nano-particles and solvent. How the optical characteristics
affect on the values of root mean square radius of gyration ${\left\langle {R_g}^2\right\rangle}^{1/2} $ need to be explored further. This also causes an important physical problem, what is the relationship between the mass and refractive index distributions? So the relationship between the optical weight root mean square radius of gyration and traditional root mean square radius need to be explored further. Due to the structures of nano-particles in dispersion, the hydrodynamic characteristics will become complex. Since the apparent hydrodynamic radius are determined by optical 
and hydrodynamic characteristics of different size nano-particles together, the 
dimensionless shape parameter $\rho =\left\langle R_{g}^{2}\right\rangle ^{1/2}/R_{app,h}$ has also lost the physical significance to judge the shape of nano-particles.

\section{CONCLUSION}

In general, the nano-particles are made through chemical reactions to form cross-linked soft particles or  
hyper-branched chains with 
a three dimensional network structure and have different density and different optical characteristics at different parts. However, our experiments reveal that the long time average data of scattered intensity can be perfectly described by the homogeneous spherical model. 
Based on the size distribution obtained using the SLS or TEM technique and the 
relation between the static and hydrodynamic radii, all the calculated and measured values of 
$g^{\left( 2\right) }\left( \tau \right) $ investigated are also consistent. Since the
the long time average scattered intensity of non-homogeneous spherical nano-particles can be described by homogeneous spherical model,
therefore the phenomenons of ergodicity happen in our experiments.
Since the root mean square radius of gyration ${\left\langle {R_g}^2\right\rangle}^{1/2} $ is an optical weight size and 
the apparent hydrodynamic radius are determined by optical 
and hydrodynamic characteristics of different size nano-particles together, the 
dimensionless shape parameter $\rho =\left\langle R_{g}^{2}\right\rangle ^{1/2}/R_{app,h}$ has also lost the physical significance to judge the shape of nano-particles.


\begin{thebibliography}{7}
\bibitem{re1}
 Zimm B H 1948 J. Chem. Phys. \textbf{16} 1099.
\bibitem{re2}
 Burchard W 1983 Adv. Polym. Sci. \textbf{48} 1.
\bibitem{re3} 
Sun Y  \textit{Different Particle Size Information Obtained From Static and Dynamic Laser Light Scattering} (Thesis, 
Simon Fraser University, 2004). 
\bibitem{re4} 
Brown W \textit{Dynamic Light Scattering: The Method and Some Applications}
(Clarendon Press, Oxford, 1993).
\bibitem{re5}
Gao J and Frisken B J 2003 Langmuir 19 5212.
\bibitem{re6} 
Berne B J and Pecora R \textit{Dynamic Light Scattering}
(Robert E. Krieger Publishing Company, Malabar, Florida, 1990).
\bibitem{re7} 
van de Hulst H C \textit{Light Scattering by Small Particles} (Dover Publications, Inc. New York, 1981).

. 

\end{thebibliography}
\end{document}